\documentclass[letter]{WileyMSP-template}
\usepackage{amsmath}

\title{Quantum critical behavior of diluted quasi-one-dimensional Ising chains}

\begin{document}
\maketitle

\author{Logan Sowadski*}
\author{Thomas Vojta*}

\begin{affiliations}
Logan Sowadski\\
Department of Materials Science \& Engineering, Missouri University of Science and Technology, Rolla, Missouri 65409, USA\\
Email Address: logansowadski@mst.edu

Dr. Thomas Vojta\\
Department of Physics, Missouri University of Science and Technology, Rolla, Missouri 65409, USA\\
Email Address: vojtat@mst.edu
\end{affiliations}


\begin{abstract}
CoNb$_2$O$_6$ is a unique magnetic material. It features bulk three-dimensional magnetic order at low temperatures, but its quantum critical behavior in a magnetic field is well described by the one-dimensional transverse-field Ising universality class. This behavior is facilitated by the structural arrangement of magnetic Co$^{2+}$ ions in nearly isolated zig-zag chains. In this work, we investigate the effect of random site dilution on the critical properties of such a quasi-one-dimensional quantum Ising system. To this end, we introduce an anisotropic site-diluted three-dimensional transverse-field Ising model. We find that site dilution leads to unconventional activated scaling behavior at the quantum phase transition. Interestingly, the critical exponents of the quantum critical point are in good agreement with those of the disordered three-dimensional transverse-field Ising universality class, despite the strong spatial anisotropy. We discuss the generality our findings as well as implications for experiments.
\end{abstract}

\section{Introduction}
\label{sec:intro}
Magnetic quantum phase transitions (QPTs) have attracted significant interest in both theory and experiment for more than two decades \cite{Sachdev_2011,VojtaM03,LRVW07,Vojta_review00}. One of the simplest theoretical models featuring a QPT is the transverse-field Ising model (TFIM) \cite{Elliott}. Despite its simplicity, experimental realizations of the transverse-field Ising QPT are surprisingly rare.  Recently, fascinating behavior has been found in the magnetic material cobalt niobate, CoNb$_2$O$_6$. This material develops three-dimensional bulk magnetic order below a temperature of about $3$K. At the lowest temperatures, it undergoes a QPT  as a function of a magnetic field applied along the crystallographic $b$-axis \cite{coldea}. An intriguing result from the experimental studies is that the critical behavior of this QPT is well described by the one-dimensional transverse-field Ising universality class, despite the three-dimensional character of the magnetic order.

\begin{figure}[tb]
    \centering
    \includegraphics[width=0.55\textwidth]{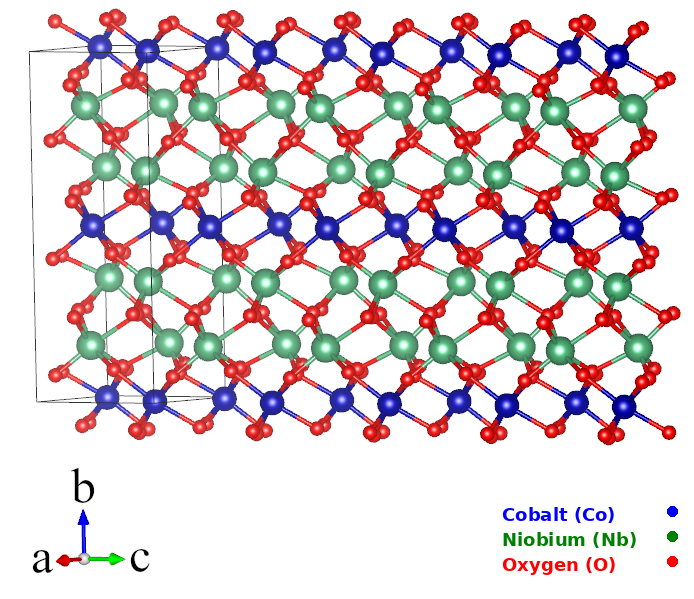}
    \caption{Crystal Structure of CoNb$_2$O$_6$. Atom positions are taken from Ref.\ \cite{matproj}.}
    \label{fig: Structure}
\end{figure}

This is caused by the fascinating crystal structure of CoNb$_2$O$_6$ \cite{Pullar}. The material forms in a columbite structure as shown in Figure \ref{fig: Structure}.
The magnetic Co$^{2+}$ ions have an effective moment of spin-$1/2$, and are embedded within a three-dimensional lattice of NbO$_6$ octahedrals. Along the $c$ direction, the Co$^{2+}$ atoms have strong magnetic interactions and form well-coupled, zig-zagging chains of magnetic sites. These one-dimensional chains are coupled to one another in the system via weak inter-chain interactions. Magnetization measurements and neutron scattering experiments have shown the exchange interactions in these weak directions to be about 10-50 times weaker than those along the strongly coupled chains \cite{coldea, thota}. Whereas these weak couplings are necessary to establish long-range order at nonzero temperatures, their influence on the critical behavior is apparently small, resulting in the one-dimensional TFIM universality class observed experimentally \cite{coldea}.

More recent investigations of cobalt niobate have shown that its complete magnetic behavior is considerably more complicated than a TFIM. In a conventional TFIM, the ferromagnetic state features two degenerate ground states with static domain walls at zero field. However, terahertz spectroscopy experiments have shown that in CoNb$_2$O$_6$, the domain wall are already mobile in the absence of a transverse magnetic field \cite{Morris_2021}. This 'quantum motion' of the domain walls is not compatible with the TFIM framework. It has therefore been proposed that even though the quantum critical behavior is well described by the one-dimensional TFIM universality class, the complete magnetic properties are better described by a twisted Kitaev chain \cite{Morris_2021}.

In this work, we investigate the effects of quenched disorder on the QPT of quasi-one-dimensional quantum magnets such as cobalt niobate. Specifically, we consider site dilution, i.e, the substitution of magnetic ions with vacancies or non-magnetic ions. Site dilution greatly challenges the formation of long-range magnetic order in the system. For the pure system without vacancies, magnetic correlations can be established with relative ease along the strongly-coupled chains of magnetic ions in the $c$ direction. A weak inter-chain coupling is then sufficient to produce three-dimensional long-range order. However, the substitution of magnetic sites with non-magnetic ones ``breaks'' the strongly-coupled chains, forcing the system to rely on the significantly weaker inter-chain interactions to reach long-range magnetic order even along the chains. Quenched disorder is known to have dramatic effects on QPT, for reviews, see, e.g., Refs.\ \cite{Vojta_2006, Vojta,vojta2019disorder}. It can cause exotic infinite-randomness scaling \cite{Fisher92,Fischer1,MMHF00,IgloiMonthus05}, Griffiths singularities \cite{ThillHuse95,rieger1996griffiths}
 and smearing \cite{vojta2003smearing,HoyosVojta08}. The problem at hand adds the additional complication of a possible dimensional crossover induced by the site dilution together with the strong spatial anisotropy.

To model this physical situation, we introduce a quasi-one-dimensional TFIM with quenched site dilution. The quantum Hamiltonian is mapped onto a four-dimensional classical Ising model with columnar disorder. This allows us to employ highly efficient classical Monte Carlo cluster algorithms,  reducing the overall numerical cost of the simulations of this model. In this way, we calculate the magnetic properties of the model and determine the quantum critical point. To gain insight into the critical behavior and the universality class of the QPT, finite-size scaling techniques are used to determine values for the tunneling, correlation length, and order parameter critical exponents. We find that this QPT features unconventional dynamical scaling behavior due to the presence of quenched disorder. We further demonstrate that the QPT belongs to the disordered three-dimensional TFIM universality class, despite the strong spatial anisotropy of the Hamiltonian, whereas the corresponding clean QPT is well described by the (clean) one-dimensional TFIM universality class.

The remainder of the paper is organized as follows. We introduce the site diluted quasi-one-dimensional TFIM in section 2. There, we also detail the quantum-to-classical mapping, the implementation of quenched disorder, and the spatial anisotropy. Section 3 describes the Monte Carlo simulations and the necessary data analysis techniques. The simulation results and critical exponents are presented in Section 4. We conclude and summarize our findings in Section 5.

\section{Model}
\label{sec:model}
\subsection{Quantum Hamiltonian}
\label{subsec:H_Q}
The magnetic-field driven QPT of pure cobalt niobate is well described by the one-dimensional TFIM. However, as discussed in Section \ref{sec:intro}, in the presence of site dilution, it is crucial to incorporate weak inter-chain couplings into the model, as magnetic long-range order is impossible otherwise, even at zero temperature. We therefore begin our study by defining the Hamiltonian of a three-dimensional anisotropic site-diluted TFIM on a cubic lattice \cite{Elliott},
\begin{equation}
H = -\bar{J}_s \sum_{\langle i,j \rangle_s} \epsilon_i \epsilon_j \sigma_i^z \sigma_j^z - \bar{J}_{\perp} \sum_{\langle i,j \rangle_{\perp}} \epsilon_i \epsilon_j \sigma_i^z \sigma_j^z - B \sum_i \epsilon_i \sigma_i^x.
\label{eq:3DTFIM}
\end{equation}
Here $\sigma_i^x$ and $\sigma_i^z$ are Pauli matrices that represent the spin-$1/2$ degree of freedom at a lattice site $i$. The first term represents interactions between nearest neighbors along the strongly-coupled chains, with an interaction strength $\bar{J_s}$. The second term accounts for nearest-neighbor interactions in the two spatial directions perpendicular to the chains. These sites are linked by the interaction strength $\bar{J}_{\perp}$. The ratio $\bar{J}_{\perp}/\bar{J}_s$ tunes the anisotropy of interactions in the system. In order to properly represent the physics of quasi-one-dimensional magnets such as cobalt niobate, $\bar{J}_{\perp}$ must be chosen to be sufficiently weak,  $\bar{J}_s \gg \bar{J}_{\perp}$. The third term represents the uniform transverse field $B$, which couples to the $x$-component of the spins, i.e.,  the Pauli matrices $\sigma_i^x$. The $\sigma_i^x$ operators can be decomposed as $\sigma_i^x = \sigma_i^+ + \sigma_i^-$, where $\sigma_i^+$ and $\sigma_i^-$ are spin-flip operators that toggle between spin-up and spin-down states. The transverse field $B$ thus introduces quantum fluctuations that can disrupt the long-range ferromagnetic order in the system when the field strength reaches a critical threshold $B_c$. This threshold is the quantum critical point which is studied in this work. Finally, the $\epsilon_{i}$ are independent quenched random variables that account for site dilution. They take the value $1$ for a magnetic site with probability $1-p$, and $0$ for a vacancy with probability $p$. Note that we do not include the more complicated interactions discussed in Ref. \cite{Morris_2021} as they are not expected to affect the critical behavior of the QPT.

\subsection{Quantum-to-classical mapping}
\label{subsec:mapping}
The  computer simulations of the three-dimensional TFIM can be simplified by mapping the quantum Hamiltonian (\ref{eq:3DTFIM}) onto an equivalent classical model which can then be studied by means of  highly efficient classical Monte Carlo cluster algorithms. Mapping the quantum Hamiltonian to a classical model involves extending the dimensionality of the system. Specifically, the thermodynamic behavior of a $d$-dimensional quantum system can be mapped onto that of a ($d+1$)-dimensional classical system \cite{Sachdev_2011}. Here, we map the three-dimensional TFIM (\ref{eq:3DTFIM}) to a four-dimensional classical Ising model on a hypercubic lattice. This system has three spatial dimensions, viz. the single strongly-coupled direction labeled by $s$ and the two weakly coupled directions labeled by $\perp$. The fourth dimension represents imaginary time $\tau$. The resulting four-dimensional classical Ising model Hamiltonian reads
\begin{equation}
    H = -J_{s} \sum_{\langle i,j\rangle_{s},\tau} \epsilon_i \epsilon_j S_{i,\tau} S_{j,\tau} - J_{\perp}\sum_{\langle i,j \rangle_{\perp},\tau} \epsilon_i \epsilon_j S_{i,\tau}S_{j,\tau}-J_\tau\sum_{i,\tau}\epsilon_iS_{i,\tau}S_{i,(\tau+1)}.
\label{eq:4DCIM}
\end{equation}
Here $i$ and $j$ denote spatial lattice positions, and $\tau$ is the imaginary time coordinate. The first term considers the couplings in the strong spatial direction, with an interaction strength $J_{s}$. The second term contains the interactions in the weak spatial directions, and the last term is the interaction in the imaginary time direction. $S_{i,\tau}$ denotes the classical Ising spin at each lattice site $(i,\tau)$.

The interactions $J_s$ and $J_{\perp}$ in the classical Hamiltonian are related to $\bar{J_s}$, $\bar{J_{\perp}}$ of the quantum model (\ref{eq:3DTFIM}). Analogously, the value of the imaginary-time interaction $J_\tau$ is determined by the transverse-field strength $B$. As we are interested in the universal properties of the phase transition, the exact values of the interactions are unimportant. We therefore fix $J_s = J_\tau = 1$ and use $J_{\perp}$ to control the spatial anisotropy of the model. We have performed test calculations for $J_\perp/J_s$ from 0.04 to 0.001. Most production calculations employ $J_\perp/J_s =0.01$, comparable to the (upper end of the) anisotropy estimates for cobalt niobate.

The phase transition is tuned via the effective temperature of the classical model, $T$. It differs from the temperature of the original quantum model, which remains at zero \cite{Sachdev_2011}. Due to the anisotropy of the classical Hamiltonian (\ref{eq:4DCIM}), we need to distinguish the system size in the strongly-coupled spatial direction, $L_s$; the size in the weakly-coupled (interchain) spatial directions, $L_\perp$; and the imaginary time size, $L_\tau$. The total number of lattice sites is thus given by $N=L_sL_\perp^2L_\tau$.

\subsection{Site dilution}
\label{subsec:dilution}
Quenched site dilution, i.e., the substitution of non-magnetic vacancies for magnetic sites, is implemented via the independent quenched random variables $\epsilon_i$. As before, they take the values $0$ (vacancy) with probability $p$ and $1$ (magnetic site) with probability $1-p$. Thus, $p$ indicates the concentration of the vacancies. As the vacancy positions are independent of (imaginary) time, the resulting disorder in the classical Hamiltonian (\ref{eq:4DCIM}) is columnar, i.e, perfectly correlated in the imaginary time direction. We expect the most interesting regime to be the weak-dilution regime.  A relatively small $p$-value allows the magnetic sites to still form sizable chains in the strongly-coupled direction. This should, in principle, permit observables to feature one-dimensional behavior, at least in a transient regime (as observed in the experiments on the clean compound \cite{coldea}).

\section{Monte Carlo Simulations}
\label{sec:simulations}
\subsection{Algorithm}
\label{subsec:algorithm}
The work reported in this paper has been performed by employing large-scale Monte Carlo simulations of the mapped classical Ising model (\ref{eq:4DCIM}). The appropriate choice of simulation algorithm is paramount for the efficiency of the simulations.

We utilize a hybrid approach, combining the Wolff cluster \cite{WOLFF} and Metropolis single-spin flip \cite{metro} algorithms. The Wolff algorithm greatly reduces the critical slowing down of the system near criticality and enables us to study larger systems at a reasonable computational cost. However, in the presence of site dilution, the Wolff algorithm alone is insufficient, as it may fail to update small, isolated spin clusters that are disconnected from the main lattice. To address this, we pair this algorithm with Metropolis single-spin updates which consider all sites, including those disconnected from the main lattice. Consequently, a full Monte Carlo sweep in our simulation consists of one Wolff cluster sweep (a number of cluster flips such that the total number of flipped spins equals the total number of lattice sites, $N = L_s L_\perp^2L_\tau$) followed by one Metropolis sweep over the lattice. This two-step approach ensures that all regions of the lattice, including isolated clusters, are adequately equilibrated, even for the large system sizes (up to almost $N=10^8$ lattice sites) we study.

\subsection{Equilibration and measurement}
\label{subsec:equilibration}
\begin{figure}[tb]
    \centering
    \includegraphics[width=0.55\textwidth]{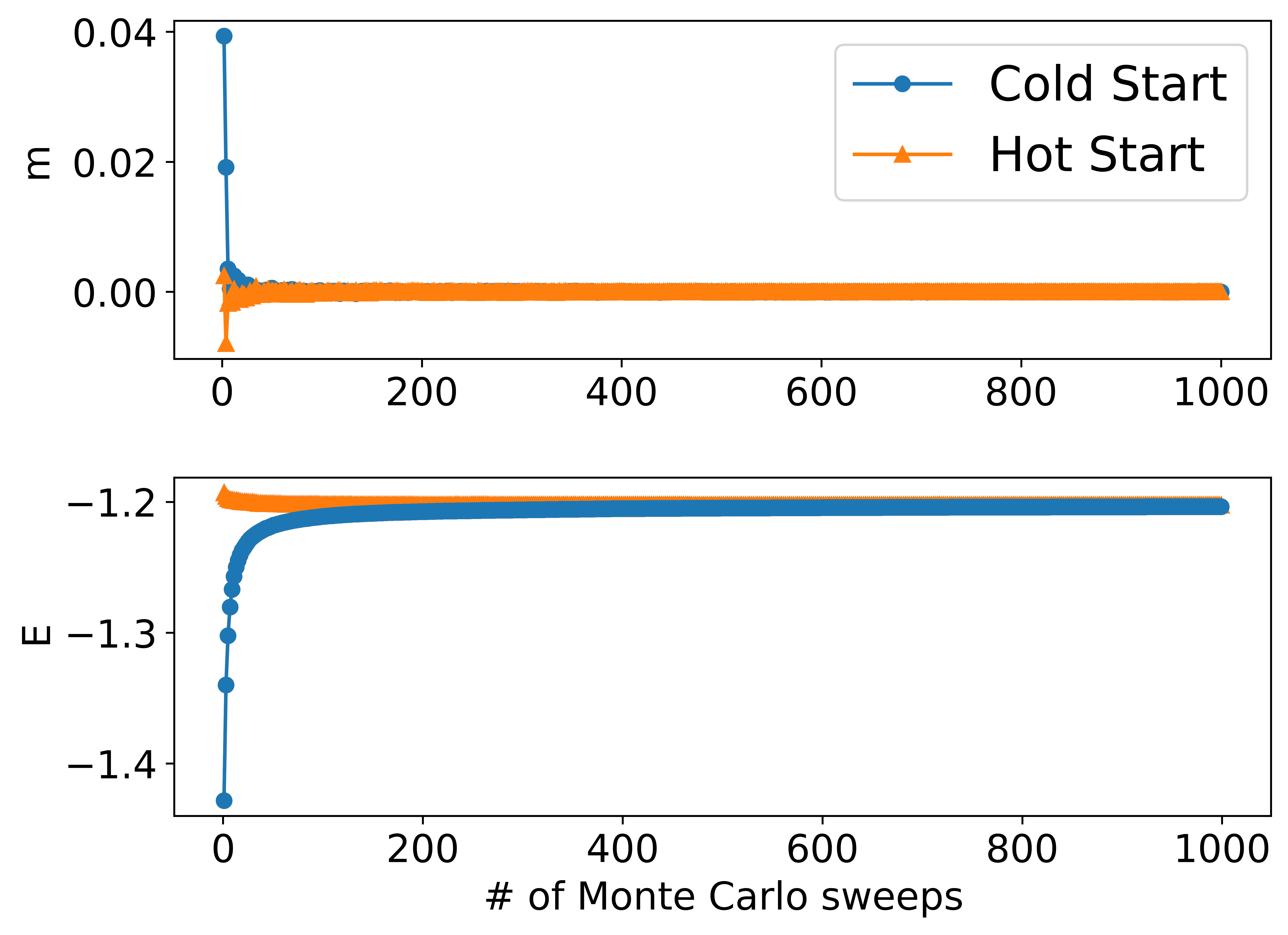}
    \caption{Equilibration test: Order parameter $m$ and energy $E$ versus Monte Carlo time (number of Monte Carlo sweeps) for a single sample of large system with optimized geometry: $L_\tau = 3080 $, $L_s = 280$, $L_\perp = 7$, comparing hot and cold initial conditions. The interactions are $J_s = J_\tau = 100J_\perp$, and the dilution is $p=0.1$. The classical temperature, $T=2.22$, is close to the phase transition.}
    \label{fig:EQ}
\end{figure}

Each simulation begins with an equilibration phase, where the system undergoes a series of sweeps until it reaches a steady state. We estimate the equilibration time by comparing runs employing hot starts (random initial spin configurations) and cold starts (all spins initially aligned). Figure \ref{fig:EQ} illustrates the equilibration process, by showing the order parameter $m$ and energy $E$ versus the number of Monte Carlo sweeps for one of our largest systems close to the phase transition.
When the values from both initial conditions agree within their statistical errors and remain stable, the system is considered equilibrated.

We observe that up to about 500 sweeps are required for the largest systems considered in this study to reach equilibrium. Based on these observations, we perform 1000 equilibration sweeps followed by 500 measurement sweeps in the Monte Carlo production runs,  with a measurement taken after each measurement sweep. The quenched disorder creates additional sample-to-sample variations of the results in our system. To suppress these disorder fluctuations, all physical quantities are averaged over $2000$ to $20,000$ independent disorder configurations for each system size. Statistical errors of the observables are obtained from the statistics of the sample-to-sample fluctuations.  Simulating a large number of disorder configurations using relatively short Monte-Carlo runs has been shown to reduce the overall statistical error for a given numerical effort \cite{BFMM98,Rotor}.

\subsection{Data analysis and sample geometry}
\label{subsec:data_analysis}

A number of physical quantities are considered in this paper to investigate the critical behavior of the system. The basic observable is the order parameter $m$, defined as,
\begin{equation}
m = \frac{1}{N_{\text{mag}}} \sum_{i,\tau} \epsilon_i S_{i,\tau}.
\label{eq:orderpar}
\end{equation}
Here,  $N_{\text{mag}} = pN = p L_sL_\perp^2L_\tau$ is the number of magnetic sites.  A robust method of determining the location of the critical point is based on calculating the disorder-averaged Binder cumulant, which is defined as
\begin{equation}
g_{\text{av}} = \left[1 - \frac{\langle | m |^4 \rangle}{3 \langle | m |^2 \rangle^2}\right]_{\text{dis}}  ~.
\label{Binder}
\end{equation}
Here, $\langle \ldots\rangle$ denotes the thermodynamic (Monte Carlo) average, whereas $[\ldots ]_{\text{dis}}$ corresponds to the average over the quenched disorder.  The Binder cumulant is a dimensionless quantity. Its finite-size scaling behavior, discussed below, is therefore particularly suited for analyzing the phase transition.

In the presence of strong spatial anisotropy, samples with $L_s = L_\perp$ are not ideal for the Monte Carlo simulations. This issue arises because the magnetic correlations decay much more slowly in the strongly coupled direction than in the weak directions. To address this, we adjust the aspect ratio of our samples by elongating the lattice in the direction of strong coupling. This transformation results in a bar geometry with $L_s > L_\perp$. This geometry makes the QPT more easily identifiable in the analysis of the Binder cumulant.  In test calculations for the clean case, $p = 0$, we use ratios $L_s/L_\perp$ between 2 and 10. 
Simulations of the diluted case, $p > 0$, benefit from an even larger $L_s/L_\perp$ ratio. In our production calculations, we employ a ratio of  $L_s/L_\perp= 40$ for systems with a vacancy concentration $p = 0.1$ and $J_\perp = 0.01$.

The ratio as $L_s/L_\perp$ is kept fixed as the system size is varied for finite-size scaling because both $L_s$ and $L_\perp$ are (spatial) lengths and are expected to have the same scale dimension.
In contrast, in a disordered quantum system, the imaginary time length $L_\tau$ must be treated as an independent parameter with a different scale dimension than the spatial sizes, $L_s$ and $L_\tau$. This comes from the fact that the disorder, which is perfectly correlated in imaginary time, but uncorrelated in space, breaks the symmetry between space- and time-like directions in the classical Hamiltonian (\ref{eq:4DCIM}).
In disordered quantum Ising systems, correlations in space and time are expected to be related by activated scaling. This means the correlation time as expected depend exponentially on the correlation length, $ \ln(\xi_\tau) \propto \xi_s^\psi $ rather than following conventional power law scaling $\xi_\tau \propto \xi_s^z$ \cite{Fisher92,Fischer1}. Here, $\psi$ is the so-called tunneling exponent which replaces the usual dynamical exponent $z$.

The expected activated scaling behavior leads to the following finite-size scaling forms of the magnetization and the Binder cumulant:
\begin{eqnarray} 
 m &=& L_s^{-\beta/\nu} \, \tilde{m}_A (tL_s^{1/\nu},L_\perp/L_s,\ln(L_{\tau})/L^\psi)~, \label{eq:orderp_ascale} \\
g_{\text{av}} &=& \qquad\quad \tilde{g}_A(tL_s^{1/\nu},L_\perp/L_s,\ln(L_\tau)/L_s^\psi)  ~. \label{eq:ascale}
\end{eqnarray}
Here  $\tilde{m}_A$ and $\tilde{g}_A$ are scaling functions, $t=(T-T_c)/T_c$ measures the distance from the critical point, $\beta$ is the order parameter exponent, and $\nu$ is the correlation length critical exponent. These activated finite-size scaling forms differ from the corresponding scaling forms for conventional dynamical scaling,
\begin{eqnarray}
 m &=& L_s^{-\beta/\nu} \, \tilde{m}_A (tL_s^{1/\nu},L_\perp/L_s, L_\tau/L_s^z)~, \label{eq:orderp_cscale} \\
 g_{\text{av}} &=& \qquad\quad \tilde{g}_A(tL_s^{1/\nu},L_\perp/L_s,L_\tau/L_s^z)~. \label{eq:cscale}
\end{eqnarray}
The value of $\psi$ (or $z$) is not known a priori and needs to be found together with the critical point in the simulations. This is equivalent to finding sample shapes ($L_\tau$ vs $L_s$) that keep the last argument of the scaling functions $\tilde{m}_A$ and $\tilde{g}_A$ constant upon system size changes.  We follow the approach outlined in Ref.\  \cite{Rotor} to determine the so called 'optimal' value of $L_\tau$ for each $L_s$. This is the value for which the ratio $L_\tau/L_s$ roughly corresponds to the ratio of correlation lengths in time and space $\xi_\tau/\xi_s$.

\begin{figure}[tb]
    \centering
    \includegraphics[width=0.5\textwidth]{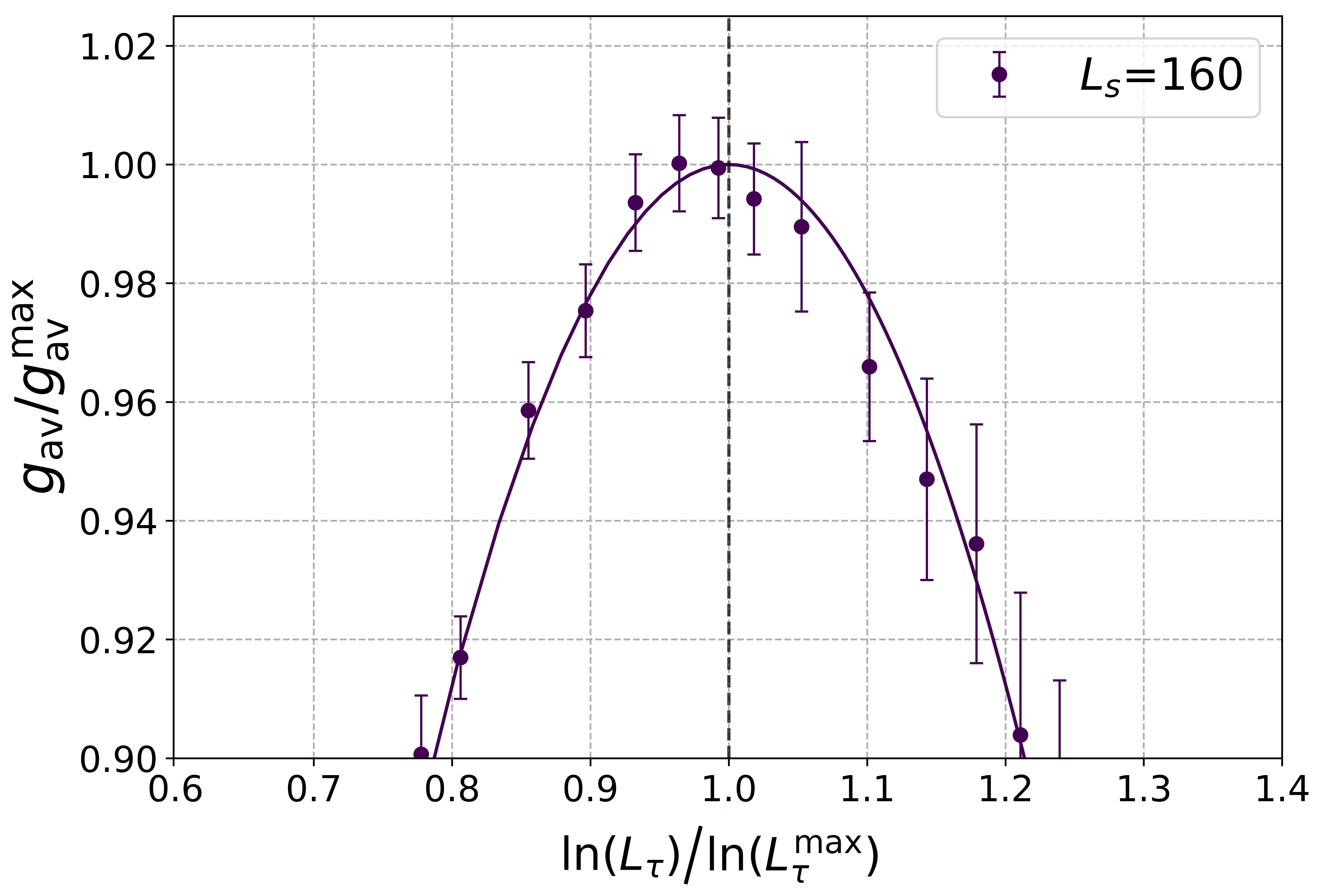}
    \caption{Normalized Binder cumulant $g_{\text{av}}/g_{\text{av}}^{max}$ at $L_s=160$, $T = 2.21$, $p = 0.10$, $J_\perp = 0.01$ vs.\ system size $L_\tau$ in imaginary time direction. Error bars indicate the uncertainty in $g_{\text{av}}$ obtained from the standard deviation over disorder realizations. Dashed vertical line indicates maximum obtained via fitting with (\ref{gavfit}).}
    \label{fig:SimpleDomes}
\end{figure}

Within this approach, the optimal value of $L_\tau$ is determined by analyzing the parabolic nature of the function $g_{\text{av}}(L_\tau)$, for constant $L$ and $T$ (see Figure \ref{fig:SimpleDomes}). 
This function has its maximum at position $L_\tau^{max}$ which indicates the optimal sample shape because the Binder cumulant is largest when the spins in the entire sample are correlated. To determine the maximum of $g_{\text{av}}(L_\tau)$, we perform a parabolic fit of the simulation results according to
\begin{equation}
g_{\text{av}}(L_\tau) = C - A(\ln(L_\tau)-\ln(L^{max}_{\tau}))^2 
\label{gavfit}
\end{equation}
where $C$, $A$ and the maximum position $\ln(L^{max}_\tau)$ are fit parameters. For the further data analysis, we then use the optimal shapes $L_\tau = L_\tau^{max}$ for each $L_s$. This fixes the last argument in the scaling functions in eqs.\ (\ref{eq:orderp_ascale}) to (\ref{eq:cscale}) and allows us to employ conventional (one-parameter) finite-size scaling in terms of $tL^{1/\nu}$.

\section{Results}
\label{sec:results}
\subsection{Finding the critical point $T_c$}
\label{subsec:findingTc}

The critical point $T_c$ is determined via an iterative approach together with the optimal system sizes $L_\tau^{max}$ in the imaginary time direction. For each spatial system size $L_s$, we begin by determining the optimal values of $L_{\tau}$ at several temperatures around the transition, following the approaches detailed in Section \ref{subsec:data_analysis}. According to the finite-size scaling forms (\ref{eq:ascale}) and (\ref{eq:cscale}), the value of $g_{\text{av}}$ for the optimal sample shape ($L_\tau = L_\tau^{max}$) is independent of $L_s$ at the critical point, $t = 0$, whereas it is expected to increase with $L_s$ in the ordered phase and to decrease with $L_s$ in the disordered phase. The resulting algorithm for finding $T_c$ and the optimal shapes is summarized in  Table \ref{tab: Steps}.  
\begin{table}[tb]
    \centering
    \caption{Algorithm for finding the critical temperature $T_c$ and the optimal sample shapes $L_\tau^{max}$, see Ref \cite{Rotor}.}
    \label{tab: Steps}
    \renewcommand{\arraystretch}{1.2} 
    \begin{tabular}{|p{14cm}|}
    \hline
    \textbf{1.} Select set of spatial system sizes $L_s$ and  $L_\perp$, with fixed ratio $L_\perp/L_s$ \\
    \hline
    \textbf{2.} Select initial set of corresponding imaginary time sizes $L_\tau$, for example $L_\tau = L_s$.\\
    \hline
    \textbf{3.} Select classical temperature range, $T = [T_-,T_+]$, bracketing the critical point. \\
    \hline
    \textbf{4.} Perform Monte Carlo simulations to compute Binder cumulant $g_{\text{av}}$ for these systems.\\
    \hline
    \textbf{5.} Increase the $L_\tau$ values and repeat from step 3. \\
    \hline
    \textbf{6. } Plot ``domes'' $g_{\text{av}}$ vs $L_{\tau}$ for each $L$ and $T$ and determine $g_{\text{av}}^{max}$ and $L_{\tau}^{max}$. \\
    \hline
    \textbf{7.} Identify $T_c$ as temperature for which the maximum value $g_{\text{av}}^{max}$ of the domes is $L$-independent.\\
    \hline
    \end{tabular}
\end{table}

\begin{figure}[tb]
    \centering
    \includegraphics[width=1.0\textwidth]{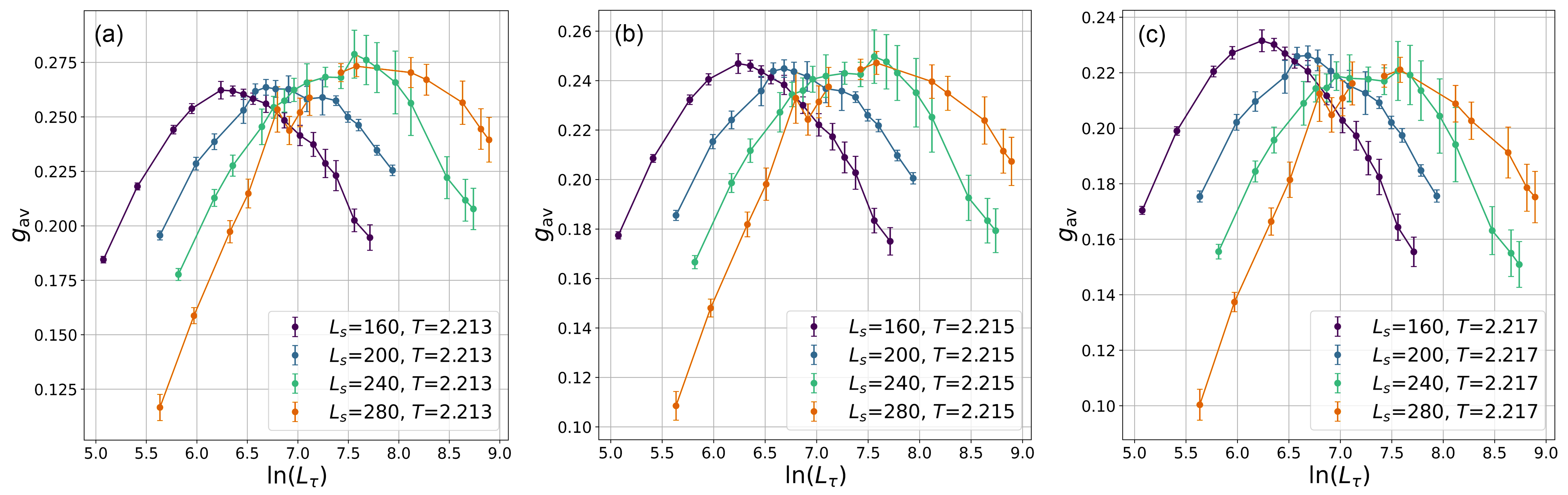}
    \caption{Binder cumulant $g_{\text{av}}$ vs. $\ln(L_\tau)$ for several $L_s$ and dilution $p=0.1$. The ratio $L_s/L_\perp$ is fixed at 40. The interaction strengths are $J_s = J_\tau = 100J_\perp$. (a) $T =2.213$. (b) $T = 2.215$. (c) $ T =2.217$.}
    \label{fig: Figure15}
\end{figure}

\begin{figure}[tb]
    \centering
    \includegraphics[width=0.55\textwidth]{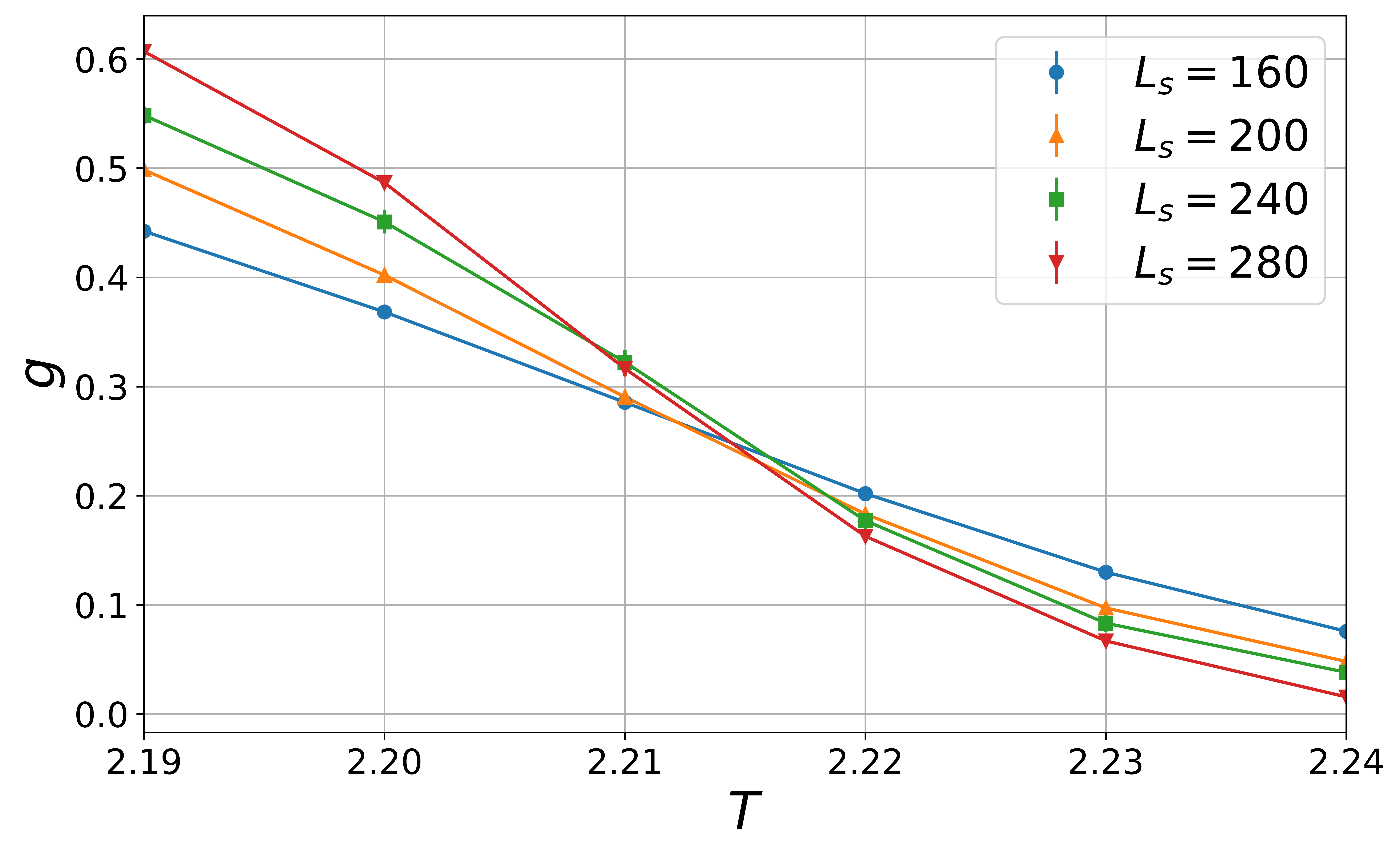}
    \caption{Binder cumulant $g_{\text{av}}$ vs.\ temperature $T$ for interaction strengths $J_s = J_\tau = 1 = 100J_{\perp}$ and the dilution is $p=0.1$. The sample geometry is given by $L_s = 40L_{\perp}$. The imaginary time size corresponds to the optimal shapes $L_{\tau} = L_{\tau}^\text{max}$ (values given in Table \ref{tab:Ltau}).  Statistical errors are about a symbol size or smaller. The lines between the data points serve as visual aids only.}
    \label{fig:BinderCrossing}
\end{figure}

The results of this analysis for the disordered quantum system with dilution $p = 0.1$ are shown in Figure \ref{fig: Figure15}. 
For $T= 2.213$ (Figure \ref{fig: Figure15}a), the parabolas shift upwards, to larger $g_{\text{av}}$, with increasing spatial system size $L_s$. This indicates that the system is in the ordered phase. For $T=2.217$, in contrast, the parabolas shift downward with increasing spatial  system size $L_s$, indicating that the system scales towards the paramagnetic phase (Figure \ref{fig: Figure15}c). At $T= 2.215$ (Figure \ref{fig: Figure15}b) the maximum value $g_{\text{av}}^{max}$  of the parabolas is approximately size-independent, indicating that the system is at the critical point. The resulting optimal values of $L_{\tau}$ at the estimated $T_c$ are determined for each spatial system size $L_s$ as the positions $L_\tau^{max}$ of the Binder cumulant maxima. They are given in Table \ref{tab:Ltau}.
\begin{table}[tb]
  \centering
  \renewcommand{\arraystretch}{1.2} 
  \setlength{\tabcolsep}{10pt}      
  \caption{Optimal imaginary time system size $L_\tau^{\text{max}}$ for each $L_s$ at the critical temperature $T = 2.215$, $p = 0.1$ and $ J = J_{\tau} = 100J_{\perp}$.}
  \label{tab:Ltau}
  \begin{tabular}{|c|c|c|c|} 
    \hline 
    $L_{s}$  & $L_\perp$ & $L_{\tau}^{max}$ & \(L_{\tau}^{max}/L_s\) \\
    \hline
    160 & 4 & 663(6)  & 4.14(3)  \\
    200 & 5 & 1158(14)& 5.79(6)  \\
    240 & 6 & 1750(33)& 7.79(13) \\
    280 & 7 &  2921(94)& 10.43(31)\\
    \hline 
  \end{tabular}
\end{table}
Further simulations are now performed by fixing $L_\tau$ at its optimal value for each $L_s$.  The value of $T_c$ can be reconfirmed by identifying the crossing of the Binder cumulant curves $g_{\text{av}}$ vs. $T$, computed for the optimal shapes, as is done in Figure \ref{fig:BinderCrossing}. This analysis yields the critical temperature for $p = 0.1$ and $J_{\perp} = 0.01$ as $T_c = 2.215(3)$.

\subsection{Critical behavior}
\label{subsec:critical}

\begin{figure}[tb]
    \centering
    \includegraphics[width=0.40\textwidth]{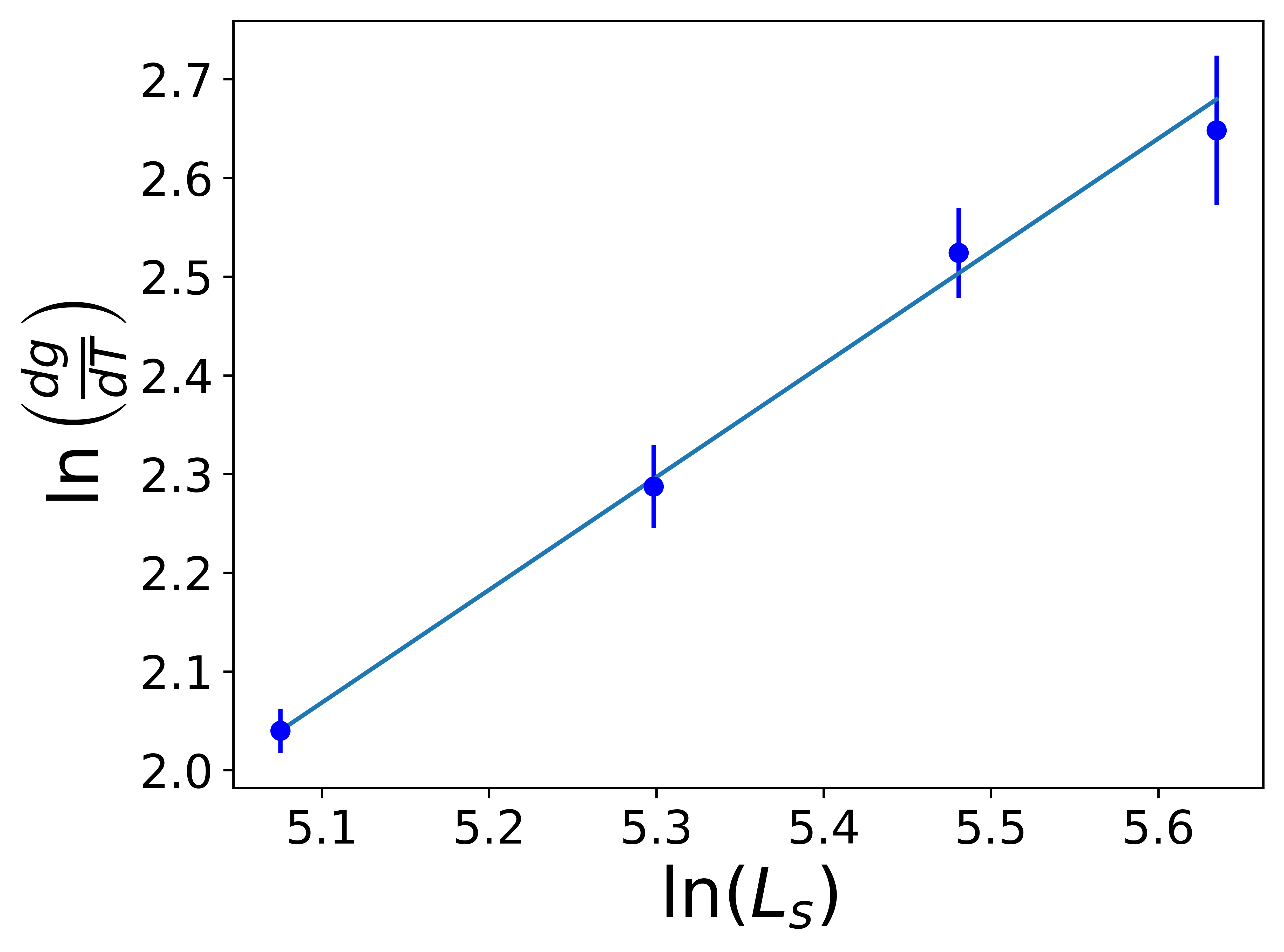}
    \caption{$\ln({dg}/{dT})$ vs. $\ln(L_s)$ at the critical point $T = T_c$ for samples of the optimal shape, $L_\tau = L_\tau^\text{max}$.  Solid line is a fit with the power law (\ref{eq:nusrelation}), yielding $\nu = 0.90(5)$.}
    \label{fig: NU}
\end{figure}

After having identified the critical point, we now turn to determining its critical behavior. The correlation length critical exponent $\nu$ can be found from the slope of the Binder cumulant curves at $T_c$. Taking the derivative of the scaling form  (\ref{eq:ascale}) with respect to $T$ yields
\begin{equation}
    \frac{dg}{dT} \propto L_s^{1/\nu} 
    \label{eq:nusrelation}
\end{equation}
which holds at the critical point $T = T_c$ for samples of the optimal shape. The derivatives ${dg}/{dT}$ can be evaluated from the simulation data by fitting each $g_{\text{av}}$ vs.\  $T$ curve with a linear dependence near the critical point.  (Errors of the derivative are estimated from the maximum and minimum slopes of Binder cumulant curves in which the data points are shifted by the statistical error of  $g_{\text{av}}$.)  To evaluate $\nu$, we plot ${dg(L_s)}/{dT}$ vs. $L_s$ in Figure \ref{fig: NU}
A fit of the data with the power law (\ref{eq:nusrelation}) is of good quality and gives $1/\nu = 1.11(6)$  which implies $ \nu = 0.90(5)$.

\begin{figure}[tb]
    \centering
    \includegraphics[width=0.40\textwidth]{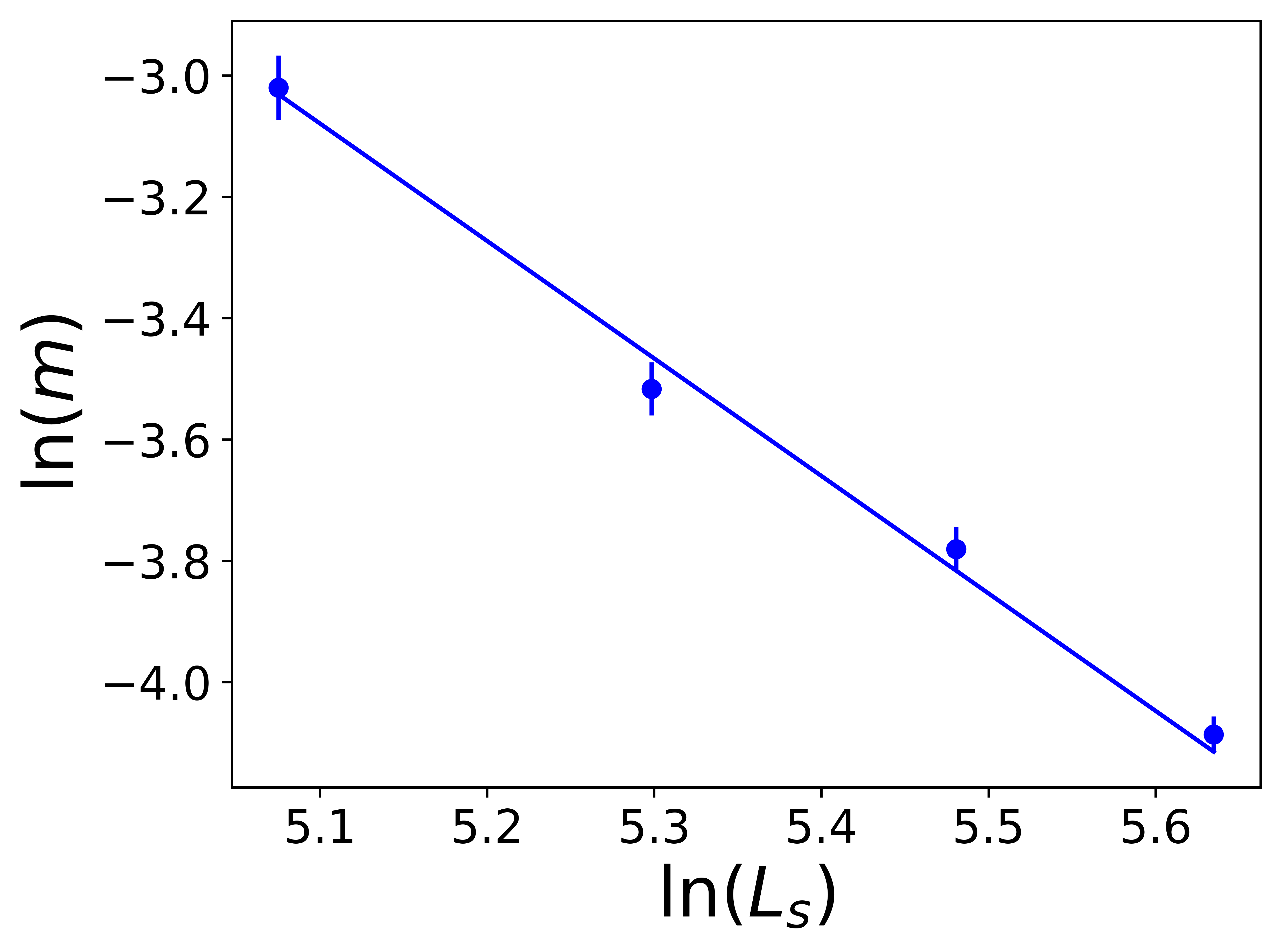}
    \caption{$\ln(m)$ vs.\  $\ln(L_s)$ at the critical point $T = T_c$ T for samples of the optimal shape, $L_\tau = L_\tau^\text{max}$.  The line is a fit with the power law (\ref{eq:orderprelation}) from which $\beta/\nu = 1.94(12)$ is extracted. Error bars indicate the statistical error of $m$, from the calculated uncertainty over disorder realizations.}
    \label{fig: BETA}
\end{figure}

The scaling behavior of the order parameter at the critical point follows from its finite-size scaling form  (\ref{eq:orderp_ascale}). At the critical point, $T = T_c$, and for samples of the optimal shape, $L_\tau = L_\tau^{max}$, this implies the relation 
\begin{equation}
    m \propto L_s^{-\beta/\nu} ~.
    \label{eq:orderprelation}
\end{equation}
Thus, we evaluate the dependence of $m$ on $L_s$ in Figure \ref{fig: BETA}.
The data follow the expected power-law relation, and a fit with eq.\  (\ref{eq:orderprelation}) yields ${\beta}/{\nu} = 1.94(12)$. Together with the value for $\nu$ found above, this
implies $\beta = 1.74(20)$.

\begin{figure}[tb]
    \centering
    \includegraphics[width=0.5\textwidth]{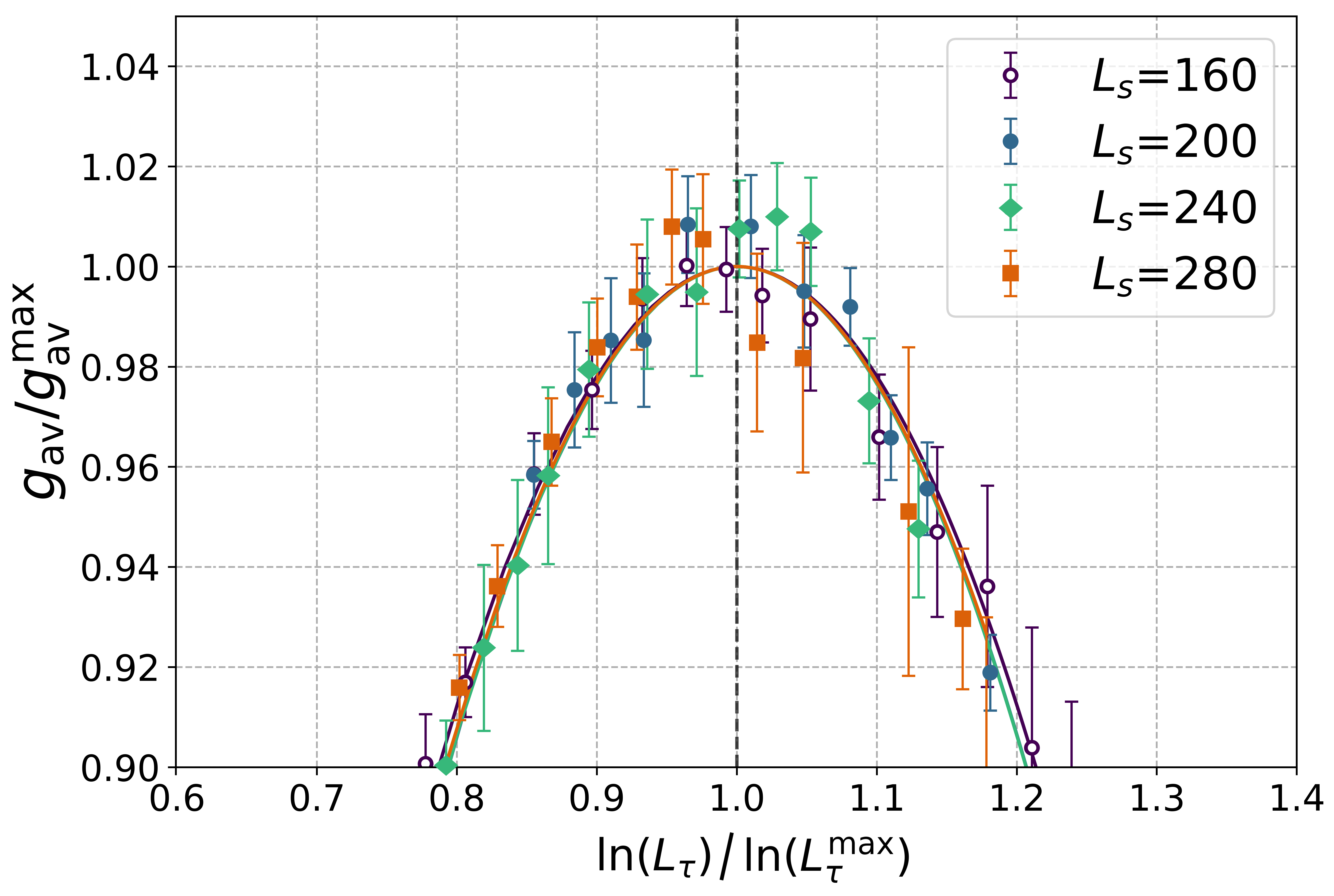}
    \caption{Scaling plot of $g_{\text{av}}/g_{\text{av}}^{max}$ vs $\ln(L_\tau)/\ln(L_\tau^{max})$ for multiple spatial system sizes $L_s$ at the critical point $T = 2.215$. The parabolas for different system sizes $L_s$ collapse onto one another.}
    \label{fig:FULLDOMES}
\end{figure}

In order to obtain a complete set of critical exponents, we now focus on the dynamical scaling. Figure \ref{fig:FULLDOMES} shows a scaling plot of the Binder cumulant $g_{\text{av}}$ at $T_c$ according to activated scaling form (\ref{eq:ascale}).
 The parabolas for different spatial system sizes collapse well within their error bars. In contrast, a scaling plot according to power-law scaling (\ref{eq:cscale}), i.e., a plot of $g_{\text{av}}$ vs.\ $L_\tau/L_\tau^{max}$,  does not lead to a good collapse, but the domes broaden with increasing $L_s$. This observation provides evidence of unconventional activated dynamical scaling behavior rather than conventional power-law dynamical scaling in this system.
 
\begin{figure}[tb]
    \centering
    \includegraphics[width=0.40\textwidth]{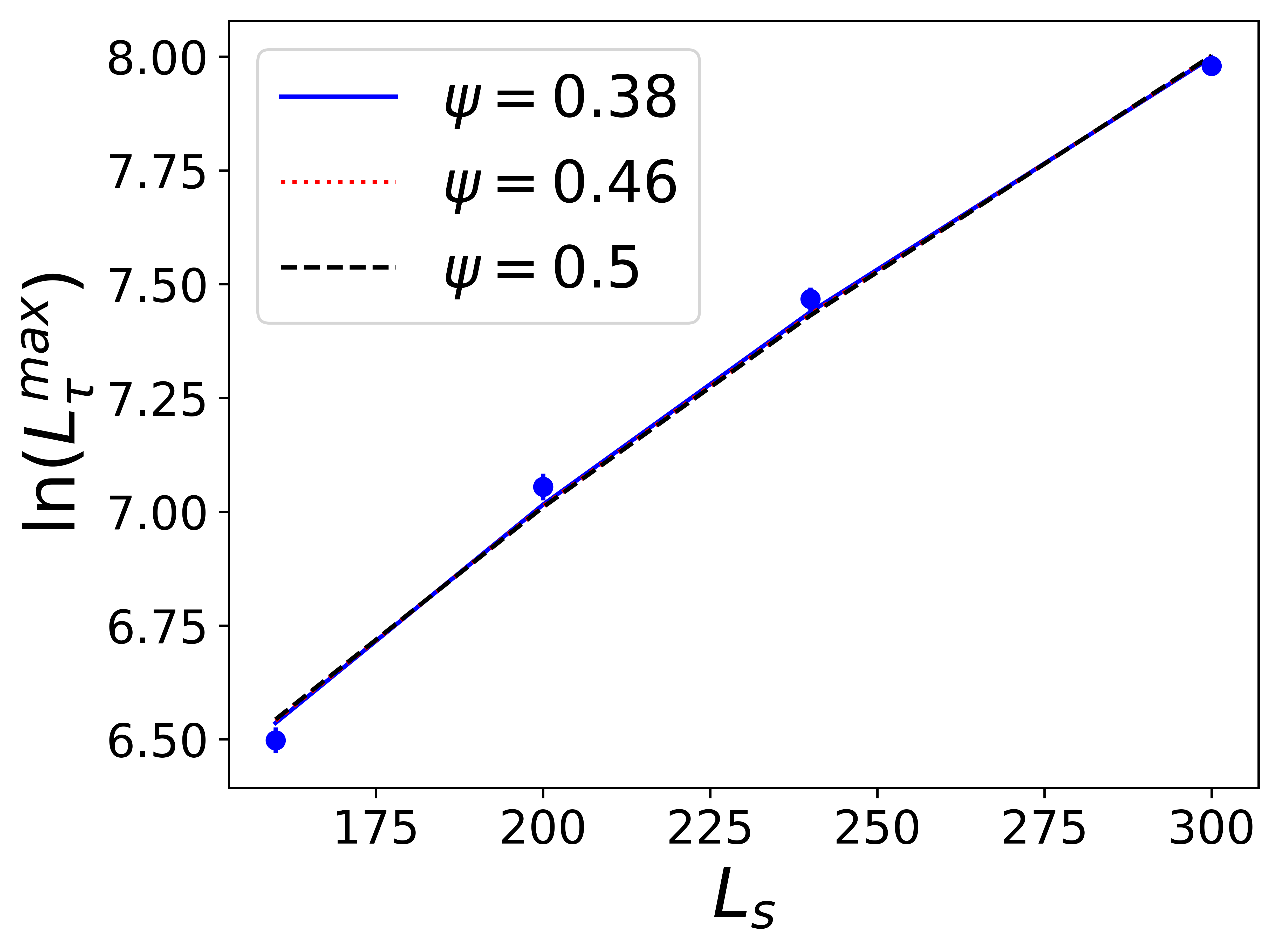}
    \caption{$\ln(L_\tau^{max})$ vs $L_s^\psi$ at the critical point $T = 2.215(3)$. The lines are fits of the data with the activated relation (\ref{eq:asrelation}). In these fits the  tunneling exponent is fixed at the indicated values. For details see text. }
    \label{fig: PSIFIGURE}
\end{figure}

The values of the optimal system size $L_\tau^{max}$ in imaginary time direction given in Table \ref{tab:Ltau} allow us to analyze the  tunneling exponent $\psi$. In the case of activated scaling, the relation between $L_\tau^{max}$  and $L_s$ is expected to take the form
\begin{equation}
    \ln(L_\tau^{max}/L_0) \propto L_s^\psi~.
\label{eq:asrelation}    
\end{equation}
Here, $L_0$ is a nonuniversal microscopic (imaginary) time scale.  Figure \ref{fig: PSIFIGURE} shows  $L_\tau^{max}$ vs.\ $L_s$ at the critical point.
Due to the presence of $L_0$, a fit of these data with eq.\ (\ref{eq:asrelation}) involves three fit parameters, viz., $\psi$, $L_0$, and the proportionality constant. This nonlinear three-parameter fit of just four data points is numerically unstable and only converges for carefully chosen initial conditions. In this case, it yields $\psi=0.33(20)$. The large error
reflects the instability of the fit. To test the robustness of this result, we have also performed two-parameter fits, fixing the  tunneling exponent $\psi$ at the values 0.38 (reported for the three-dimensional disordered contact process \cite{Vojta-CE}), 0.46 (a strong-disorder renormalization group prediction for the three-dimensional random transverse-field Ising model \cite{SDRG}),  and 0.5 (the exact value for the one-dimensional random transverse-field Ising model \cite{Fischer1}). As can be seen in the figure, the fits with these three values are practically indistinguishable.

\subsection{Discussion}
\label{subsec:discussion}
To identify the universality class of the QPT of the quasi-one-dimensional site-diluted TFIM studied in this work, we now compare our calculated critical exponents with those obtained in literature for the random TFIM in one and three space dimensions. We also compare the exponents with those of the three-dimensional disordered contact process, which is expected to belong to the same universality class as the three-dimensional random TFIM. The exponent values are presented in Table \ref{tab: Results}.
\begin{table}[tb]
    \centering
    \renewcommand{\arraystretch}{1.2} 
    \setlength{\tabcolsep}{10pt}      
    \caption{Critical exponents for the site diluted quasi-one-dimensional TFIM studied in this paper compared to the exponents of the three-dimensional disordered contact process \cite{Vojta-CE}, a strong-disorder renormalization group prediction for the three-dimensional random TFIM \cite{SDRG}, and Fisher's exact results for the one-dimensional random TFIM \cite{Fischer1}.}
    \label{tab: Results}
    \begin{tabular}{|c|c|c|c|} 
        \hline
        System & $\psi$ & $\nu$ & $\beta/\nu$ \\
        \hline
        This Work        & 0.33(20) & 0.90(5) & 1.94(12) \\
        3D Disordered Contact Process \cite{Vojta-CE} & 0.38(3)  & 0.98(6) & 1.90(4)  \\
        3D Random TFIM \cite{SDRG},   & 0.46(2)  & 0.99(2) & 1.84(2)  \\
        1D Random TFIM \cite{Fischer1} & 1/2      & 2       & $(3-\sqrt{5})/4\approx 0.191$ \\
        \hline 
    \end{tabular}
\end{table}

The table shows that the critical exponents calculated for our system are in good agreement with those found in previous work for the three-dimensional random TFIM universality class.. In contrast, they do not agree with the exponents of the one-dimensional random TFIM. More specifically, the tunneling exponent $\psi$ does not allow us to discriminate between the one-dimensional and three-dimensional random TFIM universality classes. However, our values for  $\nu$ and $\beta/\nu$ agree within their errors bars with those reported for the
 three-dimensional disordered contact process and the three-dimensional random TFIM, whereas they are far away from the corresponding values  of the one-dimensional random TFIM.

\section{Conclusions}
\label{sec:conclusions}

In this paper, we have constructed a simple model to investigate the effects of nonmagnetic vacancies on the magnetic QPT of quasi-one-dimensional quantum Ising magnets such as cobalt niobate. This model consists of a three-dimensional TFIM with a large spatial anisotropy and site dilution. To gain insight into the quantum phase transition, we have mapped the quantum Hamiltonian onto a four-dimensional classical Ising model. We have employed large-scale Monte Carlo simulations for systems with up to almost $10^8$ lattice sites to analyze the magnetic properties at the quantum critical point and to study its critical behavior.

 The results of the simulations demonstrate that the quantum phase transition in the presence of site dilution features unconventional scaling associated with an infinite-randomness fixed point. This agrees with the behavior of many other disordered QPT and with the classification of such transitions according to the effective  dimensionality of the rare regions
 \cite{VojtaSchmalian05,VojtaHoyos14}. According to this classification, rare regions in  quantum Ising magnets with undamped dynamics have an effective dimensionality of unity, leading to infinite-randomness criticality and power-law quantum Griffiths singularities. By utilizing finite-size scaling techniques, we have calculated the critical exponents $\nu$, $\beta/\nu$, and $\psi$ that completely characterize the universality class of the QPT. We have found that the critical exponents agree with those published in the literature for the disordered three-dimensional TFIM universality class, but clearly disagree with those of the one-dimensional TFIM.

This is an interesting result, particularly because one-dimensional quantum critical behavior was observed experimentally in clean undiluted cobalt niobate  \cite{coldea}. It supports the notion that a key effect of the vacancies consists in   ``breaking''  the strongly-coupled one-dimensional chains of spins into finite-size pieces. In the absence of dilution, the chains can develop magnetic long-range order by themselves (without interchain couplings), at least at zero temperature.  Therefore, the weak interchain couplings are only needed to establish coherence between the chains.
However, this behavior does not translate well to the site-diluted system because the vacancies prevent the formation of long-range magnetic order along the strongly coupled chains. Magnetic order in the diluted system therefore crucially depends on the weak interchain interactions and is intrinsically three-dimensional. Note that these arguments do not rely on the specific form of the quantum Hamiltonian. They should therefore apply to a broad class of quasi-one-dimensional quantum magnets.

We hope that this work inspires experiments on site-diluted quasi-one-dimensional Ising magnets. For example, introducing non-magnetic impurities into cobalt niobate should allow our predictions to be tested and open avenues to study the three-dimensional random TFIM universality class in experiment.

\bigskip
\textbf{Acknowledgements} \par 
The Monte Carlo  simulations reported in this paper were performed on the Pegasus, Foundry, and Mill clusters at Missouri S\&T.

\medskip

%

\bibliographystyle{unsrt}
\bibliography{add}    

\section*{APPENDIX}

In addition to the quantum Ising model (\ref{eq:3DTFIM}) discussed in the bulk of this paper, we have employed a  classical, three-dimensional, anisotropic Ising model, to study the relation between the spatial anisotropy of the interactions and the observability of one-dimensional behavior. We use the classical Hamiltonian
\begin{equation}
    H = -J_{s} \sum_{\langle i,j\rangle_{s}} \epsilon_i \epsilon_j S_i S_j - J_{\perp}\sum_{\langle i,j \rangle_{\perp}} \epsilon_i \epsilon_j S_iS_j. 
    \label{eq:ClassicalH}
\end{equation}
It takes a similar form as eq.\  (\ref{eq:4DCIM}), but without the imaginary time dimension. To test the effects of anisotropy on this system, we consider large ratios $J_s/J_{\perp}$ and analyze their effects on the magnetic correlations.

\begin{figure}[t!]
    \centering
    \includegraphics[width=0.5\textwidth]{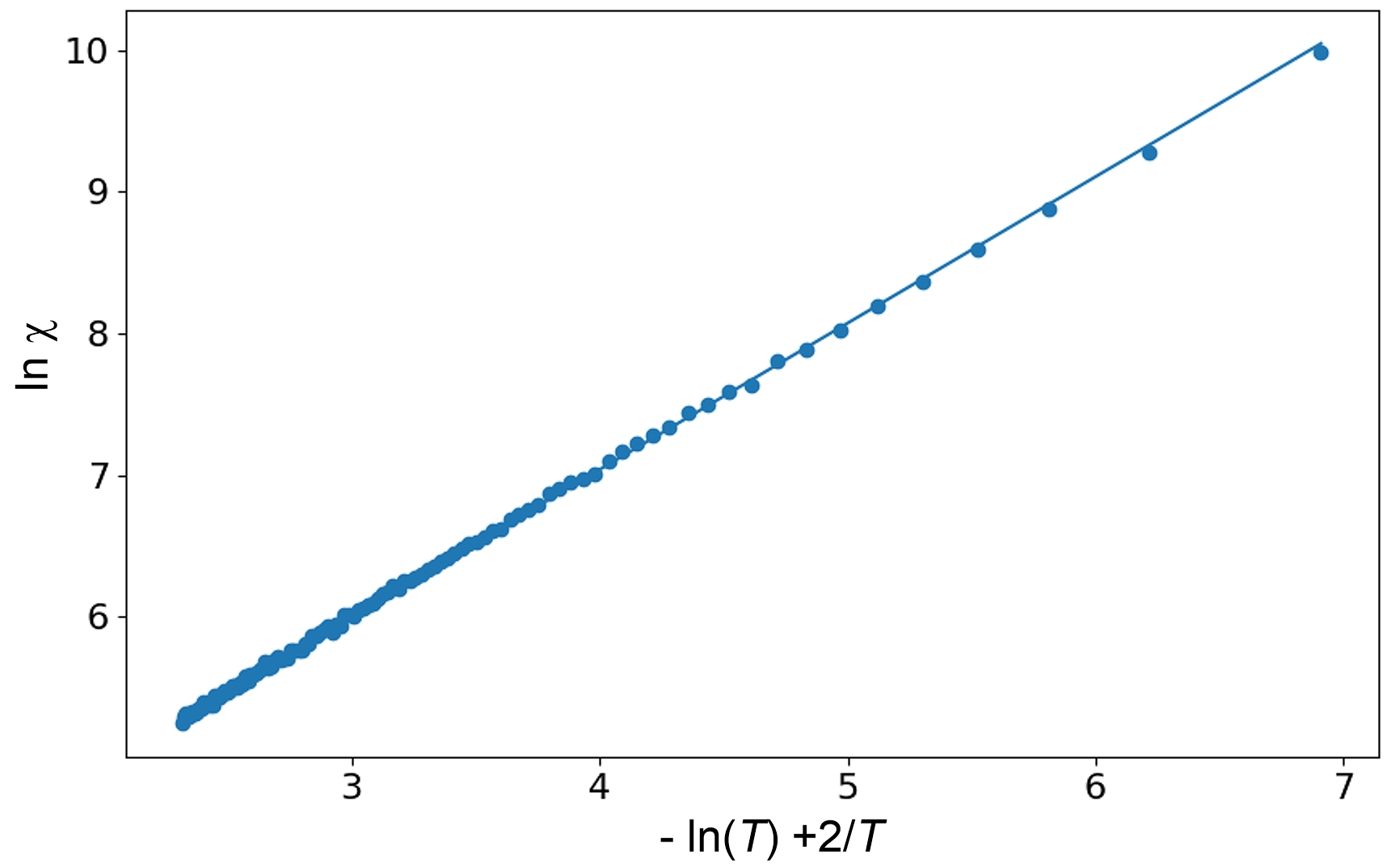}
    \caption{Magnetic susceptibility $\chi$ vs temperature $T$, plotted such that a straight line indicates agreement with one-dimensional functional form (\ref{eq:pathria}). The data
    are for the clean ($p = 0$) three-dimensional classical Ising system (\ref{eq:ClassicalH}) with high spatial interaction anisotropy $J_s = 100J_\perp$.}
    \label{fig: susceptiblity}
\end{figure}

To identify potential one-dimensional physics in this system, we use classical Monte Carlo simulations to compute the magnetic susceptibility $\chi$ is given by 
\begin{equation} 
\chi = [N_{\text{ mag}} \beta (\langle m^2 \rangle - \langle m \rangle^2)]_{\text{ dis}} 
\label{eq:suscp} 
\end{equation} 
To verify that a spatial interaction anisotropy similar to the experimental one \cite{coldea,Larsen} leads to one-dimensional behavior in the absence of dilution,  we compare the calculated susceptibility for the model (\ref{eq:ClassicalH}) for $J_s=1$ and different $J_{\perp}$ with the exact classical one-dimensional result, see e.g. Ref.\ \cite{PATHRIA}, 
\begin{equation} 
\chi \propto \frac{1}{T} \exp(\frac{2}{T}). \label{eq:pathria} 
\end{equation} 
Figure \ref{fig: susceptiblity} demonstrates that the susceptibility of a clean system with $J_{\perp} = 0.01 J_s$ is an excellent agreement with the functional form (\ref{eq:pathria}) for a one-dimensional Ising chain.

\end{document}